\documentclass[galaxies,article,accept,moreauthors,pdftex,10pt,a4paper]{mdpi}
\firstpage{1} 
\articlenumber{x}
\doinum{10.3390/------}
\pubvolume{xx}
\pubyear{2018}
\copyrightyear{2018}
\history{Received: 19 September 2017; Accepted: 19 December 2017; Published: 4 January 2018}

\Title{Optical/infrared polarised emission in X-ray binaries}

\Author{David M. Russell $^{1}$}

\AuthorNames{David M. Russell}

\address{%
$^{1}$ \quad New York University Abu Dhabi, PO Box 129188, Abu Dhabi, UAE; dave.russell@nyu.edu}

\abstract{Recently, evidence for synchrotron emission in both black hole and neutron star X-ray binaries has been mounting, from optical/infrared spectral, polarimetric, and fast timing signatures. The synchrotron emission of jets can be highly linearly polarised, depending on the configuration of the magnetic field. Optical and infrared (OIR) polarimetric observations of X-ray binaries are presented in this brief review. The OIR polarimetric signature of relativistic jets is detected at levels of $\sim 1$--10 \%, similar to AGN cores. This reveals that the magnetic geometry in the compact jets may be similar for supermassive and stellar-mass BHs. The magnetic fields near the jet base in most of these systems appear to be turbulent, variable and on average, aligned with the jet axis, although there are some exceptions. These measurements probe the physical conditions in the accretion (out)flow and demonstrate a new way of connecting inflow and outflow, using both rapid timing and polarisation. Variations in polarisation could be due to rapid changes of the ordering of the magnetic field in the emitting region, or in one case, flares from individual ejections or collisions between ejecta. It is predicted that in some cases, variable levels of X-ray polarisation from synchrotron emission originating in jets will be detected from accreting Galactic black holes with upcoming spaceborne X-ray polarimeters.}


\conference{Polarised Emission from Astrophysical Jets}

\begin{document}

\section{Introduction}
X-ray binary jets come in two main `flavours', both producing synchrotron emission -- steady, continuously launched ejecta, characterized by a flat / inverted radio spectrum, and discrete ejecta which are optically thin. The steady jets are launched during the hard X-ray state, whereas the discrete ejecta are associated with state transitions caused by changes in the structure of the accretion flow \citep[e.g.][]{fendet04}. The total jet power is uncertain, but the jet luminosity is dominated by the higher energy photons, so for the steady jets, this depends critically on the extent of the power law from the radio to higher frequencies. There is a break in this spectrum, usually in the infrared regime \citep[e.g.][and references therein]{russet13}. At higher frequencies than this jet break the spectrum is optically thin synchrotron, which originates close to the jet base in the post-acceleration plasma.

The steady jet spectrum evolves during X-ray binary outbursts. In black hole X-ray binaries (BHXBs) it is known from a few systems that the jet break shifts from the infrared down to mm frequencies as the source softens during state transitions \citep[e.g.][]{koljet15}. In the soft X-ray state, no core jet is detected. By studying the time evolution of the jet spectrum and how it relates to changes in the inflow (mass accretion rate and accretion structure), it is possible to test how the jet properties respond to accretion flow changes, which is more difficult in active galactic nuclei (AGN) because they vary on much longer timescales.

Polarisation is a powerful, yet under-used tool to probe the physical conditions in the jet plasma. Optically thin synchrotron emission can be highly polarised if the magnetic field is highly ordered. The fractional linear polarisation (FLP) is related to the ordering of the magnetic field $f$, and the electron energy distribution $p$, as $FLP_{\rm thin} = f \frac{p~+~1}{p~+~ 7/3} = f \frac{1 ~- ~\alpha_{\rm thin}}{5/3 ~ -~ \alpha_{\rm thin}}$, where $\alpha_{\rm thin}$ is the spectral index of the optically thin synchrotron emission (we define $\alpha$ such that $F_{\nu} \propto \nu^{\alpha}$). For a completely aligned magnetic (B) field, $f=1$ and the FLP can be up to $\sim 70$--80 \%, whereas for a completely tangled field, $f \sim FLP \sim 0$ \citep{rybili79,bjorbl82}.

Up until recently, very little polarimetric studies had been undertaken of the optically thin synchrotron emission in steady jets of X-ray binaries. Radio polarisation measurements are relatively common, and have revealed that steady, flat spectrum radio jets are usually polarised on levels of a few percent, whereas optically thin discrete ejecta are polarised at a higher level of up to tens of per cent \citep[e.g.][]{hannet00,gallet04,brocet07,millet08,brocet13,curret14}. The latter are likely to be largely due to B field compression in shocks, either shocks internal to the jet or shocks with the interstellar medium (ISM). In the optical regime, the accretion disc usually dominates the emission in X-ray binaries during outbursts, which is expected to be intrinsically unpolarised, with a few percent FLP expected from Thomson scattering in the disc (or companion star) photosphere \citep[e.g.][]{browet78,bochet79}. Such FLP can vary on the orbital period of the system and can even be used to constrain the system parameters such as the orbital inclination \citep{dolata89,glioet98}.

When a steady, compact radio jet is present, there is usually an OIR excess above the disc emission, that is generally attributed to the jet \citep[e.g.][and references therein]{corbfe02,kaleet13,russet13}. This is likely to be optically thin synchrotron emission from near the jet base, in the post-accelerated plasma. Polarisation measurements of this optically thin synchrotron emission can therefore probe the ordering of the B field in a region very close to the black hole ($\sim 10$--$1000 R_{\rm g}$). Models predict strong, fairly well ordered B fields close to the black hole \citep[e.g.][]{blanzn77,meieet01,vlahko04,nishet05} and more randomised, tangled or dissipated B fields at larger distances from the black hole, due to various instabilities \citep[e.g.][]{istopa94,hard04}. The field may maintain a high level of ordering over the small emission region near the jet base \citep[e.g.][]{blanko79}. On the other hand, some theoretical works have claimed that shock acceleration in the inner jets requires magnetic turbulence, and simulations have shown that instabilities amplify magnetic fields but produce turbulent B fields in the shock waves \citep[e.g.][and references therein]{zhanet17}. The geometry of the B field and how it changes downstream in the flow remain largely elusive, despite many observational and theoretical studies, especially of AGN. Polarimetric measurements of the optically thin power law in the inner jets of X-ray binaries therefore provide a powerful, underused tool to reveal the nature of the B field structure in this region, which is critical for some models and simulations of jet production. We note that at OIR wavelengths, photons do not suffer from Faraday rotation or dispersion due to its strong dependency on wavelength, which can hamper the interpretation of radio results \citep{burn66}.

Here in this concise review, examples are given of polarimetric studies of the optically thin synchrotron emission in steady jets, from optical or near-infrared (NIR) data of both black hole and neutron star X-ray binaries (NSXBs).

\section{Results}

\subsection{Examples of polarisation of BHXBs}

The BHXB GX 339--4 was observed in 2008 with ISAAC on the Very Large Telescope (VLT). Variable linear polarisation on levels of $\sim 1$--3 \% is detected in the NIR, when the steady jet dominated the emission when it was in the hard state (for details see \cite{russet11}). The position angle (PA) of polarisation was compared to the axis of the resolved radio jet of GX 339--4 \citep{gallet04}, and implies the magnetic field is approximately parallel to the jet axis. The time resolution of the polarisation measurements is less than $\sim 60$ sec, and any variability of the FLP on timescales shorter than that, if present, could be damping the measured polarisation observed on these timescales probed.

During the rapid, bright flaring episode of V404 Cyg, on 23 June 2015, optical time-resolved polarisation was obtained with the Telescopio Nazionale Galileo (TNG) on La Palma \citep{shahet16}. 2-sec exposures were made in r$^{\prime}$-band, for two hours. The intrinsic polarisation light curve was determined by subtracting the interstellar quiescent polarisation level. Variable intrinsic polarisation was observed, varying between FLP $=$ 4.5 and 3.5 \% in 20 minutes; the PA was fairly constant at $\sim 171$ deg. Very Long Baseline Interferometry radio data during the same outburst revealed the resolved radio jet, which had a changing PA of between $\sim -30$ and $+5$ degrees E of N \citep{millet18}. The optical PA implies that the electric-field vector near the base of the jet is parallel to the jet axis, and the B-field is orthogonal to the jet axis. We interpret this as evidence for magnetic field compression in shocks within the jet, resulting in a partially ordered transverse field \citep{shahet16}.

A very bright X-ray flare of V404 Cyg was observed with INTEGRAL JEM X--1, reaching about 40 Crab. A prominent radio flare at 16 GHz was observed with the Arcminute Microkelvin Imager Large Array (AMI-LA), peaking just a few hours after the optical polarisation flare. The optical polarisation flare occured during the initial stages of the rise of the radio flux. The 16 GHz and 5 GHz flares \citep{fendet18} peak at 2 h and 4 h respectively after the optical polarisation flare. The interpretation is that the jet base is first seen in polarised light at optical wavelengths, and then the radio flares, which arise from emission along the jet downstream. The three flares describe a classically evolving synchrotron flare from an ejection. It appears we are witnessing an ejection evolving from optical to radio wavelengths, over a frequency range spanning 4.5 orders of magnitude.

In a similar result, \cite{lipuet16} presented time-resolved optical polarisation of V404 Cyg using the MASTER network of telescopes during the 2015 outburst. Two flares of optical polarisation were discovered, where changes in FLP of $\sim 4$--6 \% were seen on hour-timescales. These were observed when the optical flux was faint, before an optical flare, at a similar stage of the evolution of the flares of V404 Cyg as the TNG polarisation flare. These two results, from TNG and MASTER polarimetric data, may represent the first flares of optical polarisation detected from \emph{discrete jet ejections} in an X-ray binary.

Intrinsic optical or NIR polarisation, with likely a steady jet origin, has also been reported from some other BHXBs; XTE J1118+480, XTE J1550--564, GRO J1655--40, Cyg X--1, Swift J1357.2--0933 and A0620--00 \citep{schuet04,dubuch06,russfe08,chatet11,russsh14,russet16}.

\subsection{Examples of polarisation of NSXBs}

NIR spectropolarimetry of the persistent `Z-sources' (NSXBs) Sco X--1 and Cyg X--2 revealed an excess of FLP at the longest wavelengths ($\sim 2 \mu$m) \citep{shahet08}. NIR (photometric) polarimetry was performed on Sco X--1 on several dates, which showed variable FLP in this source \citep{russfe08}. While the optical FLP (from \cite{schuet04}) was consistent with the expected level from interstellar dust, the NIR had a transient excess above the dust model. All these detections of polarisation are stronger at low frequencies, which is expected from synchrotron emission because synchrotron becomes more dominant at lower frequencies \citep[e.g.][]{wangwa14}, whereas this is not expected for the other sources of polarisation such as scattering due to interstellar dust or in the disc atmosphere. The results imply a predominantly tangled, likely variable magnetic field near the jet base, similar to what was found for the steady jets in BHXBs.

A multiwavelength campaign was performed on Cyg X--2 in 2010. Time-resolved NIR polarisation observations were made with the Long-slit Intermediate Resolution Infrared Spectrograph (LIRIS) on the William Herschel Telescope (WHT) for three hours, simultaneously with X-ray data from \emph{Swift} and the Rossi X-ray Timing Explorer (RXTE) and radio data from the Westerbork Synthesis Radio Telescope (WSRT) \citep{russet18}. From preliminary results, the NIR $K_{\rm S}$-band FLP appears to be varying on short (minute) timescales between $\sim 0.3$ to 1.0 \%, whereas the PA seems to remain constant. The PA is consistent with the B field being aligned with the direction of the jet axis, which was recently resolved in radio images \citep{spenet13}. These results imply that although the field is likely to be largely tangled, when there is a slight alignment of the field, it is preferentially along the axis of the radio jet. This is similar to the results of GX 339--4 above. From preliminary results, there appears to be no clear relation between the FLP and the NIR flux or the X-ray flux, but further tests can be done to search for inflow--outflow links (e.g. looking for lags, or correlations between the flux variability and the FLP).

Intrinsic optical FLP has been discovered from the NSXB, 4U 0614+091 (\cite{baglet14b}; see also \cite{baglet17}). Here, the observed FLP $\sim 3$ \% implies the jet synchrotron contribution may be polarised at a level up to $\sim 20$ \% (the disc dominates the optical emission). Polarimetric studies of other NSXBs have so far yielded upper limits to any intrinsic polarisation from synchrotron emission \citep[e.g. Aql X--1, Her X--1, Cen X--4, PSR J1023+0038;][]{charet80,schuet04,russfe08,baglet14a,baglet16,baglet17}.

\section{Conclusions}

From the studies made so far, it is clear that NIR--optical synchrotron emission from jets in X-ray binaries is polarised. The results so far suggest that near the jet base the magnetic field is probably:

\begin{itemize}[leftmargin=*,labelsep=5.8mm]
\item  generally turbulent (only partially ordered) and rapidly changing,
\item  this seems to be similar to AGN jet cores \citep[e.g.][]{homali06,lopeet14},
\item  parallel to the jet axis (except in V404 Cyg and Cyg X--1 which are perpendicular -- which may be due to shocks compressing the B field).
\end{itemize}

The open questions in this field are:
\begin{itemize}[leftmargin=*,labelsep=5.8mm]
\item What are the timing properties of the variable polarisation?
\item Does polarisation correlate with anything in the inflow?
\item What drives the magnetic field changes?
\end{itemize}

More data and more models are needed to explain the observations. In addition, since FLP is detected from the optically thin synchrotron emission in the steady jets, it is worth considering the extent of the optically thin power law to higher frequencies. If the power law extends to X-ray or $\gamma$-ray energies, polarisation could be detected at these energies \citep[e.g.][]{jouret12,russsh14} with new and upcoming experiments such as PoGOLite / PoGO+ \citep{chauet16}, X-Calibur \citep{guoet13}, XIPE \citep{soffet13} and IXPE \citep{weiset16}.

\vspace{6pt} 


\acknowledgments{DMR would like to thank Tariq Shahbaz, Maria Cristina Baglio, Rob Fender, Elena Gallo and James Miller-Jones for fruitful discussions on the works mentioned in this review.}


\conflictsofinterest{The authors declare no conflict of interest.} 

\reftitle{References}




\end{document}